\documentstyle[11pt,paspconf,epsf]{article}

\begin{document}

\title{BeppoSAX Observations of the BL Lac Object Mkn~501 in an Intermediate State}

\author{E. Pian, E. Palazzi}
\affil{ITESRE, C.N.R., Via Gobetti 101, I-40129 Bologna, Italy}
\author{L. Chiappetti}
\affil{IFCTR, C.N.R., Via Bassini 15, I-20133 Milan, Italy}
\author{L. Maraschi, F. Tavecchio, G. Ghisellini, G. Tagliaferri}
\affil{Osservatorio Astronomico di Brera, Via Brera 28, I-20121
Milan, Italy}
\author{G. Fossati}
\affil{SISSA/ISAS, Via Beirut 2-4, I-34014 Miramare (Trieste), Italy}
\author{A. Treves}
\affil{Dip. di Fisica, Universit\`a di Como, Via 
Lucini 3, I-22100 Como, Italy}
\author{C. M. Urry}
\affil{STScI, 3700 San Martin Drive, 
Baltimore, MD 21218}
\author{G. Vacanti}
\affil{ESA ESTEC, Space Science Department, Astrophysics Division, 
Postbus 299, NL-2200 AG Noordwijk, The Netherlands}

\begin{abstract}

The BL Lac object Mkn~501 was observed with the BeppoSAX satellite at three epochs in
April-May 1998, simultaneously with the Whipple and HEGRA Cherenkov telescopes.  The
X-ray spectrum is well detected up to 70 keV and it exhibits, at all epochs, a
continuous curvature, which is here modeled with three power-laws of increasingly
steeper index at larger energies.  In the $\nu F_\nu$ representation the spectrum
exhibits a peak at $\sim20$ keV, which is interpreted as the maximum of the
synchrotron emission.  This implies that the synchrotron peak energy has lowered by
an order of magnitude with respect to the powerful X-ray outburst observed in April
1997.  The simultaneous TeV flux was comparable to the lowest levels observed for
Mkn~501, possibly suggesting that the peak of the inverse Compton radiation had also
shifted toward lower energies.

\end{abstract}


\section{Introduction}

Mkn~501 is one of the closest ($z$=0.034) BL Lacertae objects, and one of the
brightest at all wavelengths.  It is one of the only four extragalactic sources
detected so far at TeV energies [1,2]. Based on observations prior to 1997, its
spectral energy distribution ($\nu F_\nu$) resembles that of BL Lac objects selected
at X-ray energies, having a peak in the EUV-soft X-ray energy band.  Accordingly, the
2-10~keV spectra observed for this source were relatively steep, with energy spectral
indices $\alpha$ larger than unity ($F_\nu \propto \nu^{-\alpha}$), meaning the power
output peaks below this band [3-6]. BeppoSAX observations of Mkn~501 in April 1997
revealed a completely new behavior. The spectra showed that at that epoch the
synchrotron component peaked at 100 keV or higher energies, implying a shift of at
least two orders of magnitude of the peak energy with respect to the quiescent state
[7].  Correspondingly the source was extremely bright in the TeV band and exhibited
rapid flares [8,9].  We have reobserved Mkn~501 with BeppoSAX at three epochs in
April-May 1998, simultaneously with ground-based optical and TeV Cherenkov telescopes
(Whipple and HEGRA).  We will illustrate here the preliminary results of the spectral
analysis performed on the three average spectra and will briefly discuss the
correlated X-ray and TeV variability.

\section{Observations, analysis and results}

Mkn~501 was pointed by the BeppoSAX Narrow Field Instruments on 28, 29 April
and 1 May 1998 for $\sim$10 hours in each occasion. Data reduction was done
following standard methods (see e.g., [10,11]).
LECS data have been considered in the
range 0.1-4 keV, 
MECS data in
the range 1.8-10.5 keV, HPGSPC data in the range 6-30
keV, and PDS data in the range 13-70 keV.  For all epochs, fits to all data
either with a single or a broken power-law are unacceptable.  Therefore,
joint fits with a broken power-law to the LECS and MECS
spectra on one hand and to the MECS, HPGSPC and PDS spectra on the other have been
performed.
The fit parameters, obtained by fixing the
value of the hydrogen column density to the Galactic value ($1.73 \times
10^{20}$ cm$^{-2}$, [12]), are reported in Table~1. The $\chi^2$ values
associated with these fits are all very close to unity.  The spectral
steepening at the lower and higher break energies is $\sim$0.3 and $\sim$0.4,
respectively, at all epochs. The data suggest that the actual spectrum is 
continuously curved.

\begin{table}
\caption{Fit Parameters for BeppoSAX Spectra of Mkn~501 in 1998.}
\label{tbl-1}
\begin{center}\small
\begin{tabular}{rccccc}
Epoch & $\alpha_1$ & $E_{break,1}$ & $\alpha_2$ & $E_{break,2}$ &
$\alpha_3$ \\
      &            & keV           &            & keV           &  \\
\tableline
&&&&\\
April 28 & $0.50 \pm 0.04$ & $1.2 \pm 0.4$ & $0.81 \pm 0.05$ & $17 \pm 5$ &
$1.20 \pm 0.05$ \\
April 29 & $0.48 \pm 0.04$ & $1.3 \pm 0.4$ & $0.83 \pm 0.05$ & $17 \pm 5$ &
$1.15 \pm 0.05$ \\
May 01 & $0.62 \pm 0.04$ & $1.9 \pm 0.4$ & $1.00 \pm 0.05$ & $23 \pm 5$ &
$1.45 \pm 0.05$ \\
\end{tabular}
\end{center}
\end{table}

\section{Discussion}

The X-ray flux of Mkn~501 in the 2-10 keV range observed in April-May 1998 was
averagely at the same level as that seen on 7 April 1997, namely the lowest
state observed during that campaign [7], but brighter than ever previously
observed (see Fig. 1).  
We still observe
that the synchrotron peak is located at a very high energy, $\sim$20 keV,
which is unprecedented for any other blazar, although an order of
magnitude lower than seen in April 1997 in this source. This has
allowed BeppoSAX to fully resolve the peak of this emission component.  
The spectrum has an almost constant slope at the softer energies, and it steepens
with decreasing flux at the highest energies (Fig. 1).
It is noticeable that, although the flux and the spectrum up to $\sim$20 keV
in the present observations were very similar to those of 7 April 1997, 
the PDS spectrum is considerably steeper. 
The TeV flux measured in this period by the Whipple
and HEGRA telescopes was rather small (comparable to the first detection level,
[13]), and definitely lower than
observed in 1997 at the beginning of the simultaneous X-ray and TeV
outburst.  Since the TeV emission of Mkn~501 is produced through inverse
Compton scattering off the same relativistic particles which radiate via the
synchrotron mechanism, we are led to conclude that, analogously to the
synchrotron radiation peak, also the inverse Compton peak has shifted toward
the lower energies.  
A more quantitative discussion of the correlated X-ray and TeV
variability should await a specific model and is deferred to a future paper.

\begin{figure}
\plotone{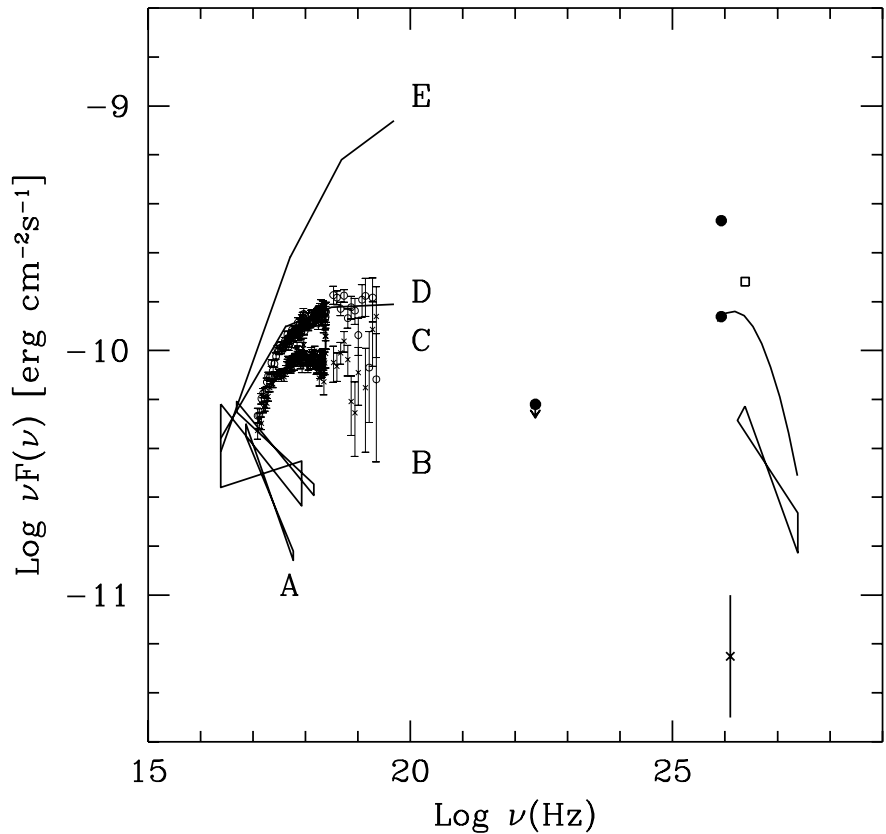}

\caption{X-ray-to-TeV energy distributions of Mkn~501.  X-ray spectral fits in low
state (A) are collected from the literature (see [7] for references).  Best-fit to
the BeppoSAX spectra on 1997 April 7 and 16 are labelled with D and E, respectively. 
The X-ray data points represent the unfolded spectra of
1998 April 29 (C, open circles) and May 1 (B, crosses).  Whipple and HEGRA TeV data
and EGRET upper limit nearly simultaneous to 7, 16 April 1997 and April-May 1998 are
indicated as open square, filled circles, and cross, respectively
[8,9,14,15]. The TeV spectral fits are from HEGRA (power-law, 15-20 Mar 1997, [9])
and from Whipple (parabolic law, Feb-Jun 1997, [16]).}

\end{figure}

\acknowledgments

We thank H. Krawczynski and J. Quinn for providing information on
the preliminary results of TeV observations in 1998.


\begin{references}

\reference [1] Quinn, J., et al. 1996, ApJ, 456, L83
\reference [2] Bradbury, S. M., et al. 1997, A\&A, 320, L5
\reference [3] Sambruna, R. M., et al. 1994, ApJS, 95, 371
\reference [4] Comastri, A., Fossati, G., Ghisellini, G., \& Molendi, S. 1997,
ApJ, 480, 534
\reference [5] Worrall, D. M., \& Wilkes, B. J. 1990, ApJ, 360, 396
\reference [6] Kubo, H., et al. 1998, ApJ, 504, 693
\reference [7] Pian, E., et al. 1998, ApJ, 492, L17
\reference [8] Catanese, M., et al. 1997, ApJ, 487, L143
\reference [9] Aharonian, F., et al. 1997, A\&A, 327, L5
\reference [10] Giommi, P., et al. 1998, A\&A, 333, L5
\reference [11] Chiappetti, L., et al. 1998, Nuc.  Phys. B (Proc. Suppl.)
69/1-3, 340
\reference [12] Elvis, M., Lockman, F. J., \& Wilkes, B. J. 1989, AJ, 97, 777
\reference [13] Quinn, J., et al. 1996, ApJ, 456 L83
\reference [14] Quinn, J. 1998, private communication
\reference [15] Krawczynski, H. 1998, private communication
\reference [16] Samuelson, F. W., et al. 1998, ApJ, 501, L17


\end{references}
\end{document}